\documentclass[prb,twocolumn,showpacs,amsmath,amssymb,superscriptaddress]{revtex4}
\pdfoutput=1

\usepackage{graphics}
\usepackage{dcolumn}
\usepackage{bm}
\usepackage{epsfig}
\usepackage{amsbsy}
\usepackage{amstext}
\usepackage[T1]{fontenc}
\usepackage[latin9]{inputenc}
\usepackage{color}

\newcommand{\be}{\begin{equation}}
\newcommand{\ee}{\end{equation}}
\newcommand{\bea}{\begin{eqnarray}}
\newcommand{\eea}{\end{eqnarray}}

\newcommand{\veps}{\varepsilon}
\newcommand{\vk}{{\boldsymbol k}}

\definecolor{green}{rgb}{0,0.75,0.3}

\begin{document}

\title{Entanglement entropy and entanglement spectrum 
 of triplet topological superconductors}
\author{T. P. Oliveira}
\email{tharnier@fisica.ufsc.br}
\affiliation{
Centro de F\'\i sica das Interac\c c\~oes Fundamentais,
Instituto Superior T\'ecnico, Universidade de Lisboa,
Av. Rovisco Pais, 1049-001 Lisboa, Portugal 
}
\author{P. Ribeiro}
\email{pribeiro@pks.mpg.de}
\affiliation{
Max Planck Institute for the Physics of Complex Systems, 
N\"othnitzer Str. 38, 01187 Dresden, Germany.
}
\affiliation{
Max Planck Institute for Chemical Physics of Solids, 
N\"othnitzer Str. 40, D-01187 Dresden, Germany,
}
\affiliation{
Centro de F\'\i sica das Interac\c c\~oes Fundamentais,
Instituto Superior T\'ecnico, Universidade de Lisboa,
Av. Rovisco Pais, 1049-001 Lisboa, Portugal 
}

\author{P. D. Sacramento}
\email{pdss@cfif.ist.utl.pt}
\affiliation{
Centro de F\'\i sica das Interac\c c\~oes Fundamentais,
Instituto Superior T\'ecnico, Universidade de Lisboa,
Av. Rovisco Pais, 1049-001 Lisboa, Portugal 
}

\affiliation{Beijing Computational Science Research Center, Beijing 100084, China}

\date{\today}

\global\long\def\ket#1{\left| #1\right\rangle }

\global\long\def\bra#1{\left\langle #1 \right|}

\global\long\def\kket#1{\left\Vert #1\right\rangle }

\global\long\def\bbra#1{\left\langle #1\right\Vert }

\global\long\def\braket#1#2{\left\langle #1\right. \left| #2 \right\rangle }

\global\long\def\bbrakket#1#2{\left\langle #1\right. \left\Vert #2\right\rangle }

\global\long\def\av#1{\left\langle #1 \right\rangle }

\global\long\def\tr{\text{Tr}}

\global\long\def\im{\text{Im}}

\global\long\def\re{\text{Re}}

\global\long\def\sgn{\text{sgn}}

\global\long\def\Det{\text{Det}}

\global\long\def\pd{\partial}

\global\long\def\abs#1{\left|#1\right|}

\global\long\def\bs#1{\boldsymbol{#1}}

\begin{abstract}
We analyse the entanglement entropy properties of a two-dimensional p-wave superconductor with Rashba spin-orbit coupling, which displays a rich phase-space that supports non-trivial topological phases, as the chemical potential
and the Zeeman term are varied. We show that the entanglement entropy and its derivatives clearly signal
the topological transitions and provides a sensible signature of each topological phase. 
We separately analyse the contributions to the entanglement entropy that are proportional to 
or independent of the perimeter of the system, as a function of the Hamiltonian coupling constants 
and the geometry of the subsystem. We conclude that both contributions are generically non-universal 
depending on the specific Hamiltonian parameters and on the particular geometry. 
Nevertheless, contributions to the entanglement entropy due to different kinds of boundaries or edges simply add up. 
We also observe a relationship between a topological contribution to the entanglement entropy in a 
half- cylinder geometry and the number of edge states, and that the entanglement spectrum has robust modes 
associated with each edge state.
\end{abstract}


\maketitle
\section{Introduction}
Topological phases of matter are characterised by global entanglement and correlations. 
Due to their non-local nature, topologically ordered phases are robust to local perturbations and have
therefore received wide attention in the context of error-free quantum computation and quantum information processing.

Although a full classification of topological phases is far from being achieved, very important steps have been given in this direction, in particular in the context of gapped non-interacting systems where a complete classification has been put forward \cite{LudwigNJP}.  
For those, the characterisation and detection of topological phases may be achieved by certain topological invariants associated with the filled energy bands. In the case of certain insulator classes such invariants can be associated with direct physical response functions such as a quantised Hall conductivity.  

A manifestation of topologically ordered phases is the appearance of gapless symmetry-protected edge modes at the interface between topologically distinct phases.

On the experimental side an increasing activity has been seen in the search for new the topological insulators and topological superconductor materials. Various experimental signatures have been proposed and considerable experimental evidence has by now been found\cite{Hasan,Qi,Alicea,Konig,Hsieh,Xia,Mourik,Das,Deng,Fan,Yazdani,Veldhorst,Williams,Rokhinson,Pablo,Schnyder,Ojanen,us3}.
 
The absence of a local order parameter and the impossibility to use Landau symmetry breaking arguments turn topological phases into one of the most successful examples where entanglement measures, such as the entanglement entropy, bring new insights that could be hardly achieved by other more traditional methods.  
The use of quantum information concepts in condensed matter physics has already been extensively considered with entanglement measures being used to probe properties of many body states \cite{amico} and to detect quantum phase transition \cite{Gu}. Kosterlitz-Thouless transitions, laking a local order parameter,  were successfully detected calculating the fidelity susceptibility of the $XXZ$ spin chain \cite{yang76,chen77}. Many other systems were also studied such as the one-dimensional Hubbard model \cite{gu77,campus78}, spin-$1/2$ particles on a torus \cite{hamma77},  the toric code model and the quantum eight-vertex model \cite{abasto78}, the spin honeycomb Kitaev model \cite{yang78,abasto79,zhao80,trebst,wang10}, as well as other spin systems \cite{castelnovo,eriksson}.

In the thermodynamic limit, the ground state of a local Hamiltonian having a finite energy gap is characterised by short range correlations of local observables. 
For such systems, the ground state entanglement entropy of a subsystem $A$:  $S_A = \tr \rho_A \ln \rho_A $ (with $\rho_A$ the reduced density matrix of the subsystem)  follows the so-called area law, i.e. it is proportional to the size of the boundary of $A$ in the limit of asymptotically large systems.  Nonetheless sub-leading corrections are expected.  For a two dimensional system with perimeter $P$  the area law \cite{eisertrmp} translates to: 
\begin{eqnarray}
S_A = \xi_A P -\gamma_A + \cdots  \ .
\label{eq:S_A}
\end{eqnarray}
Here $\xi_A$ is a non-universal constant term, $\gamma_A$ contains the non-extensive contributions and $\cdots$ 
denote other contributions that vanish as $L\to \infty $.  
In a gapped system, where correlations are short ranged,
it is expected that the entanglement entropy between two regions should follow the area law (the boundary
between the two sub regions); if a phase has topological order, where some long range entanglement is
expected, then there should be an entropy reduction with respect to the area law of a non-topological gapped
system.

The non-extensive corrections have been shown to encode subtle effects due to the presence or absence of topological and/or long range order. 
Some contributions to $\gamma_A$ are universal, in the sense that they are robust to transformations that do not close the gap, and, in
some cases, can be assigned to a given phase of mater.  
However, this information can be masked by non-universal terms dependent on the particular geometry of $A$.  

A prescription to extract the topological content of the entanglement entropy was given by Kitaev and Preskil \cite{kitaev96} and Levin and Wen \cite{levin96}. In their proposals they engineer particular ways to extract a universal topological  contribution, here denoted as $\gamma_\text{Topo}$, from the non-extensive terms.  This quantity, dubbed topological entanglement entropy (TEE), has been proposed to characterise certain topological phases in relation to the quantum dimension $D$ of the system:  $\gamma_\text{Topo}=\ln(D)$ \cite{kitaev96,levin96}.
Since then, $\gamma_\text{Topo}$ has been used as a signature of topological order in several systems as, for instance, in frustrated quantum dimer models and in the Kitaev honeycomb model \cite{misguich,hamma77}, as well as applied to detect topological order in spin liquid states \cite{balents,depenbrock}. 
A rich set of different behaviours have been found, in particular for gapped systems that spontaneously break a discrete symmetry where the correction to the area law has been observed to receive other contributions that may add up to the topological ones. This corrections are negative and are given by the logarithm of the number of degenerate groundstates \cite{balents2}.
For systems with gapped quasiparticle-like excitations but with gapless collective modes resulting from a spontaneous breaking of a continuous symmetry there are also negative corrections that diverge logarithmically as the system size grows \cite{kallin}.

For systems described by an Hamiltonian that is quadratic in the electronic operators, such as 
 topological insulators or superconductors (with an externally fixed superconducting phase),  
$\gamma_{\text{Topo}} $ vanishes. Therefore, even for systems that cannot be adiabatically connected to trivial 
band insulators, there is no topological order in the sense of Ref. \onlinecite{WenAP}. Nevertheless for a generic domain $A$,  
$\gamma_A$ is non-zero and depends on the particular domain geometry. 
This has been observed for instance in Ref.\onlinecite{Haas}  for a $p_x+ip_y$ {\em spinless} superconductor for which $\gamma_A$ 
is proportional to the number of corners of the partition. 
Moreover, even though the system has both a trivial and a nontrivial topological phases with quantum dimension $D=2$, 
$\gamma_{\text{Topo}}$ vanishes in both and does not distinguish between these phases.
A similar situation was identified in the context of Kitaev's model where for a cylinder geometry an extra contribution 
to the entanglement entropy ($\ln \sqrt{2}$) was identified \cite{Yao}.

In this work we calculate the entanglement entropy of a $p$-wave superconductor where the presence of Rashba spin orbit
coupling and a Zeeman term allow for a phase diagram that spans several topological nontrivial phases. 
We analyse both the extensive and non-extensive contributions to $\gamma_A$ in different geometries: a subsystem of an infinite system and a half of a cylinder. 
We also show that even though the TEE vanishes everywhere, a suitable choice of geometry enables the identification
of the various topological phases, through a relation to the number of edge states, in agreement with other
signatures such as the Chern and winding numbers. 
In section II we present the model studied and discuss its properties. 
In section III we present the formalism to calculate the entanglement entropy and the entanglement spectrum of a quadratic
system. In section IV we present the results. In subsection A we analise the extensive contribution to the
entanglement entropy and recover the overall characteristics of the phase diagram calculating the magnetization derivative
of the entanglement entropy. In subsection B we consider the non-extensive contributions of a subregion of an
infinite system and compare the results as a function of the shape of the various subregions and show that,
even if $\gamma_\text{Topo}$ vanishes for an asymptotically large subregion, finite size effects are considerable and can be seen to clearly signal phase transitions.
In subsection C we analise the case of a cylinder geometry and show that the non-extensive contribution to the
entanglement entropy can be related to the number of edge states.  
We conclude with section V.

\section{Triplet topological superconductor}

\begin{figure}[b]
\includegraphics[width=1.00\columnwidth]{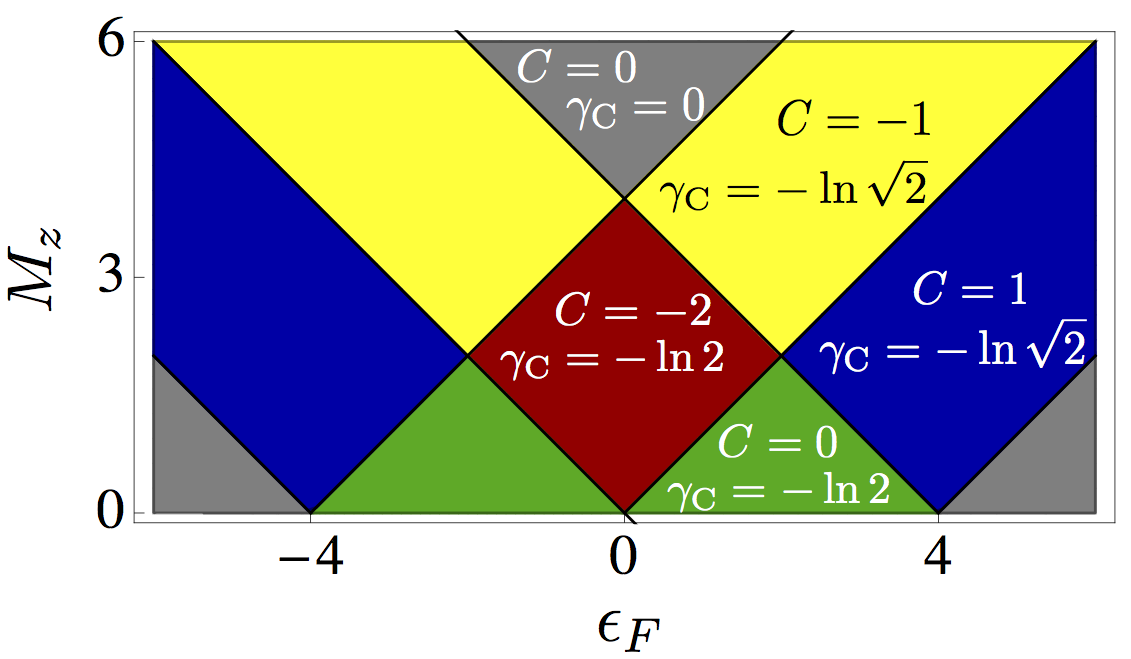}
\caption{\label{fig1}
(Color online) Topological phases and their Chern numbers, $C$, as a function of chemical potential, $\veps_{F}$, and magnetization, $M_{z}$.
$\gamma_\text{C}$ is the topological entanglement entropy on a cylinder geometry (see text).
The phases with $C=0$ and $\gamma_\text{C}=0$ are topologically trivial gapped phases.
}
\end{figure}

Superconductivity with non-trivial topology may be obtained in several different ways \cite{Qi}.
However, non-centrosymmetric superconductors such as CePt$_3$Si \cite{Bauer_2004}or the more recently discovered Li2 Pd$_x$ Pt$_{3-x}$ B \cite{Togano_2004}, where s and p-wave paring may coexist, constitute rather natural candidates from an experimental viewpoint.   
In this work we consider a model,  recently proposed in Ref. \onlinecite{sato}.
that includes a  Rashba-type spin-orbit coupling leading to a admixture of s- and p-wave pairing and a Zeeman splitting terms that breaks time reversal symmetry. 

For simplicity we consider only a triplet paring term with $p$-wave symmetry. 
In the presence of a time-reversal breaking Zeeman term and
Rashba spin-orbit coupling the phase diagram displays various transitions between topological phases characterized
by different Chern numbers, as studied before in Refs. \onlinecite{sato,us3,entmod}.
The Hamiltonian is given by
\begin{eqnarray}
\hat H = \frac 1 2\sum_\vk \bs C_{\bs k}^\dagger 
\bs H(\vk)
\bs C_{\bs k}
\label{bdg1}
\end{eqnarray}
where 
\begin{eqnarray}
\bs H(\vk) = \left(\begin{array}{cc}
\bs H_0(\vk) & \bs \Delta(\vk) \\
\bs \Delta^{\dagger}(\vk) & -\bs H_0^T(-\vk) \end{array}\right),
\end{eqnarray}
$\bs C_{\bs k}=\left(c_{\bs k\uparrow},c_{\bs k\downarrow},c_{-\bs k\uparrow}^{\dagger},c_{-\bs k\downarrow}^{\dagger}\right)^T$
and $\boldsymbol{k}$ is the wave vector in the $xy$ plane (the lattice constant is set to unity).  
The normal state Hamiltonian is given by $\bs H_0(\boldsymbol{k}) = \epsilon_\vk\sigma_0 - M_z\sigma_z + \bs H_R(\vk)$ where $\epsilon_{\boldsymbol{k}}=-2 t (\cos k_x + \cos k_y )-\epsilon_F$, with $t$ the hopping amplitude and $\epsilon_F$ the chemical potential, is the kinetic part;  $M_z$ is the Zeeman splitting field (in units of energy); and  $\bs H_R (\vk)= \boldsymbol{s} \cdot \boldsymbol{\sigma}$ is the Rashba spin-orbit term with $\boldsymbol{s} =\alpha (\sin k_y,-\sin k_x, 0)$. The Pauli matrices $\sigma_x,\sigma_y,\sigma_z$  act on the spin sector, and $\sigma_0$ is the identity.

We consider a unitary pairing contribution that reads 
$\bs \Delta = i\left( {\boldsymbol d}\cdot {\boldsymbol\sigma} \right) \sigma_y$
with the vector $\boldsymbol{d}=(d_x,d_y,d_z)$ specifying the particular $p$-wave superconducting pairing. 
As the spin-orbit term breaks parity, a singlet pairing contribution is in principle also allowed but will 
not be considered here, for simplicity.
Furthermore, we concentrate in the strong spin-orbit limit where the spin-orbit coupling is 
expected to be aligned with the paring vector 
$\boldsymbol{d}.\boldsymbol{s}= \left|\left| \boldsymbol{s} \right|\right|  
\left|\left| \boldsymbol{d} \right|\right|  \, $ \cite{Sigrist2}. If the spin-orbit coupling is not
strong, weak-pairing case, the phase diagram is less rich \cite{us3}.
Even though in each topological phase the spin-orbit coupling may be turned off without affecting the
topology, its presence, together with the time-reversal symmetry breaking Zeeman term, leads to a
nonvanishing anomalous Hall effect and a finite Hall conductivity that may be used to obtain information
about the topological phases \cite{us3}.

In the absence of the Zeeman field ($M_z=0$) time-reversal symmetry is preserved and the system belongs to the symmetry class DIII. For this class the topological invariant belongs to $\mathbb{Z}_2$ \cite{Ludwig,LudwigAIP,LudwigNJP}.
In the presence of a Zeeman term, time-reversal-symmetry (TRS) is broken and the system belongs to the symmetry class D (the TRS operator $\mathcal{T}$ is such that $\mathcal{T}^2 = -1 $). The topological invariant that characterizes this class phase is the first Chern number $C$, and the system is said to be a $\mathbb{Z}$~topological superconductor. Fig. (\ref{fig1}) shows the phase diagram, labeled by the Chern number values of each phase, as a function of chemical potential, $\veps_{F}$, and Zeeman field, $M_{z}$.

\section{Entanglement entropy and Entanglement Hamiltonian}

For quadratic Hamiltonians the entanglement entropy can be obtained calculating the eigenvalues of the single particle correlation matrix defined entirely in the subregion $A$ \cite{Peschel_2003,peschel1,peschel2}. In the following we briefly recall this procedure for a generic superconducting system and introduce some notation used in the subsequent sections. We use $\bar A$ to denote the complement of $A$. 

A quadratic many-body Hamiltonian can be written in the form $H  =  1/2 \bs C^{\dagger}\bs H\bs C $ where $\bs{H=}\bs H^{\dagger}$ is the single-body Hamiltonian and  $\boldsymbol{C}=\left\{ c_{1},...,c_{N},c_{1}^{\dagger},....,c_{N}^{\dagger}\right\} ^{T}$ a column vector of annihilation and creation operators  of $N$ fermionic modes. 
A thermal density matrix of the composite system $A+\bar A$, is thus given by $\rho=e^{-\frac{1}{2}\bs C^{\dagger}\bs{\Omega}\bs C}/Z$, 
with $\bs{\Omega} = \beta \bs{H}$, $\beta$ the inverse temperature and 
$Z=\tr{e^{-\frac{1}{2}\bs C^{\dagger}\bs{\Omega}\bs C} } =\left[ \text{det} \left( 1 + e^{-\bs{\Omega}} \right)\right]^{1/2}$.
For a electronic system whose density matrix is of the form just described, 
the reduced density matrix of any subsystem $A$ is itself of the quadratic form 
$\rho_A = e^{ -1/2 \, \bs C_A ^\dagger \bs\Omega_A \bs C_A} /Z_A$ (with $C_A$ a vector of annihilation 
and creation operators restricted to subsystem $A$). The matrix $\bs\Omega_A$ can be most simply obtained from the correlations matrix $\bs \chi_{i,j} = \av{ \bs C_i \bs C_j^\dagger }$ as \cite{Peschel_2003}:
\begin{eqnarray}
\bs{\Omega}_{A} & = & -\ln\left[\bs{\chi}_{A}^{-1}-1\right]
\end{eqnarray}
and $Z_{A} = \sqrt{\det\left(\bs{\chi}_{A}^{-1}\right)}$.  $\bs{\chi}_{A; i,j}$ is obtained from the global correlation matrix $\bs\chi$ by restricting the indices $i$, $j$ to label degrees of freedom of subsystem $A$ only. 
As a result of this simplification arising only for quadratic models, the entanglement entropy of a subsystem is given by
\begin{eqnarray}
S\left[\rho_{A}\right] 
 & = & -\frac{1}{2} \sum_{\alpha}\left( 1-\lambda_{\alpha}\right) \ln
\left( 1-\lambda_{\alpha}\right) + \lambda_{\alpha} \ln 
\lambda_{\alpha} \label{eq:S_0}
\end{eqnarray}
where $\lambda_{\alpha}$ are the eigenvalues of $\bs{\chi}_{A}$.

The entanglement spectrum of a subsystem  introduced by Li and Haldane \cite{Li_2008} is defined as set of eigenvalues of the logarithm of the reduced density matrix: $\Omega _A =  -\ln\rho_{A} + c$, up to an overall additive normalization constant $c$. The many body operator $\Omega _A$ can be interpreted as an effective Hamiltonian if the subsystem $A$ was taken to be in a Gibbs state at temperature $T=1$. 
The entanglement spectrum was shown to contain information about states along the boundary
between the two subsystems, including excited states \cite{Bernevig1,Poilblanc,Bernevig2,lukasz}.
When the topological phases result from inversion symmetry \cite{Turner} it was shown to
provide a more robust signature of the topology than the edge states.
Its usefulness is also evident in the analysis of the spectral flow of the
entanglement spectrum, in particular the trace index, which may be related to changes in the Chern
number in topological insulators \cite{Hughes_2010,Alexandradinata_2011}.

For quadratic systems it is natural to define the so called  "entanglement Hamiltonian" \cite{Li_2008}  as 
\begin{eqnarray}
\Omega_A & = & -\ln\rho_{A} - \ln Z_{A}=\frac{1}{2}\bs C^{\dagger}\bs{\Omega}_{A}\bs C
\end{eqnarray}
here the constant $- \ln Z_{A}$ was chosen in order for the spectrum of $\Omega_A $ to be particle-hole symmetric.
The eigenvalues of $\Omega_A$ can be written as 
\begin{eqnarray}
\Lambda_{\bs n} & = & \sum_{\alpha: \varepsilon_\alpha>0} (n_\alpha-1/2) \varepsilon_\alpha 
\end{eqnarray}
where $\varepsilon_\alpha$'s are the eigenvalues of $\bs{\Omega}_{A}$ and the sum runs over non-negative $\varepsilon_\alpha$. The vector $\bs n$ labels the eigenvectors of $\Omega_A$ with its $n_\alpha = 0,1$ being the occupation number of mode $\alpha$. As $\bs{\Omega}_{A}$ is the single particle operator corresponding to $\Omega_A$ it  is usually referred  as  single particle Hamiltonian and set of $\varepsilon_\alpha$'s as the single particle entanglement spectrum  \cite{Turner,Hughes_2010, Alexandradinata_2011}.

Noting that $\varepsilon_\alpha = - \ln( \lambda^{-1}_\alpha -1)$ Eq.(\ref{eq:S_0}) can also be written 
\begin{eqnarray}
S\left[\rho_{A}\right] & = & -\int d\omega\,\nu\left(\omega\right)n_{f}\left(\omega\right)\ln n_{f}\left(\omega\right)
\label{eq:S}
\end{eqnarray}
where  $\nu\left(\omega\right)  =  \sum_{\alpha}\delta\left(\omega-\varepsilon_{\alpha}\right)$ is the particle-hole symmetric $\left[ \nu\left(\omega\right)=\nu\left(-\omega\right)  \right]$ density of states of  single-particle entanglement Hamiltonian $\bs{\Omega}_{A}$ and 
\begin{eqnarray}
n_{f}\left(\omega\right) & = & \left[1+e^{\beta \omega}\right]^{-1}
\end{eqnarray}
is the Fermi function taken in Eq.(\ref{eq:S}) to have unit temperature $\beta=1$. In this form the interpretation of Eq.(\ref{eq:S}) is particular transparent, it is simply given by the sum of the entropies of the individual single particle states weighted a Fermi-statistics.

\begin{figure}[b]
\includegraphics[width=1.00\columnwidth]{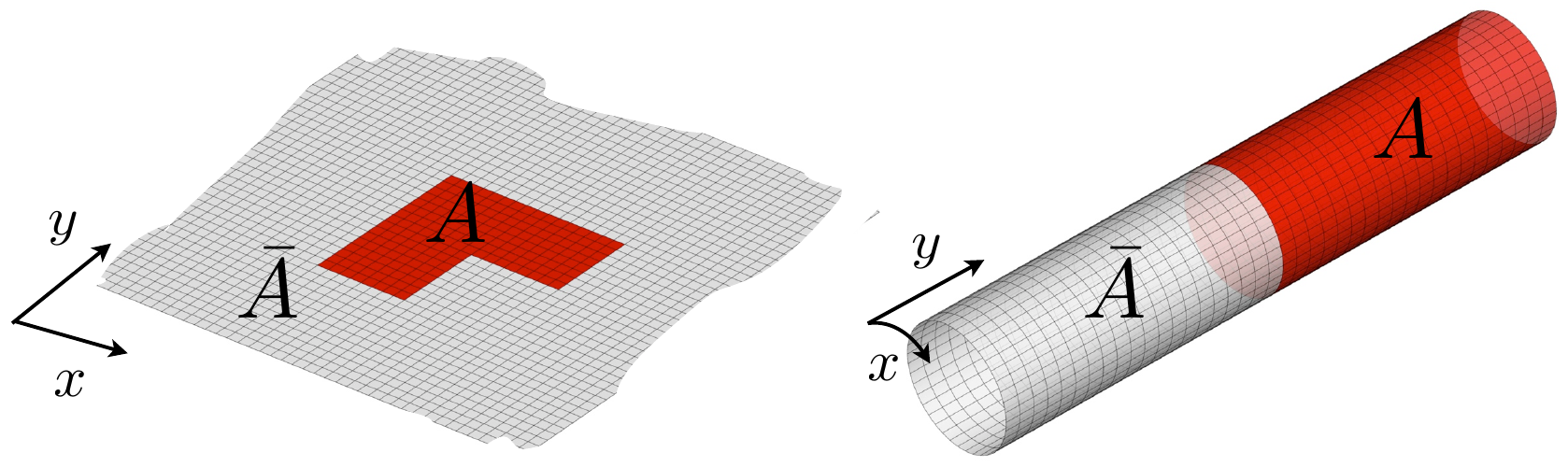}
\caption{\label{fig:geom}
Domain $A$ and its complement $\bar A$ for the two considered geometries: Left - subregion of a infinite system; Right - half of a section of a finite cylinder.  
}
\end{figure}

\begin{figure}[h]

\includegraphics[width=1.0\columnwidth]{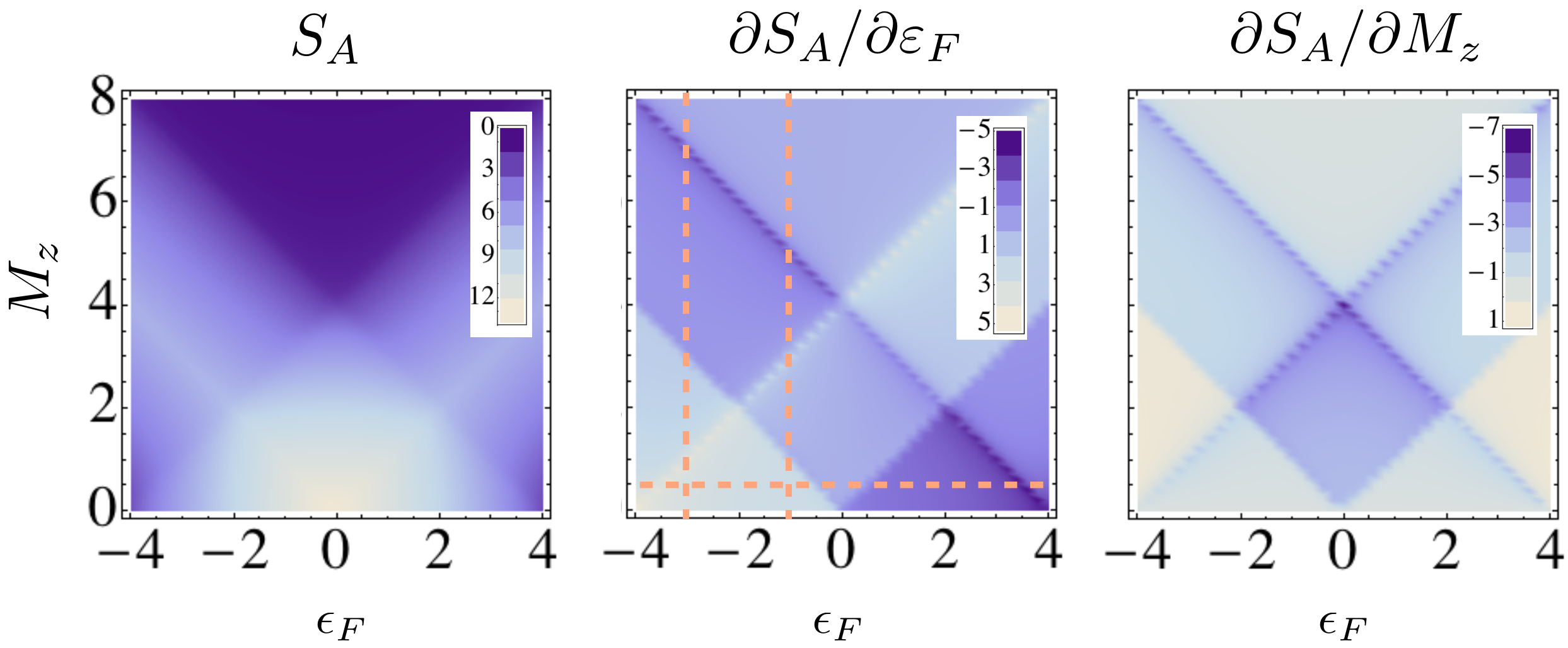}
\caption{\label{fig2}
Left panel: Local entanglement entropy of a squared $L\times L$ system with $L=6$ as a function of
chemical potential and magnetization. Middle and right panels: derivatives of the entanglement entropy with respect to the chemical potential and the Zeeman term, respectively. The orange lines correspond to specific cuts in the phase diagram that are considered in the following. 
}
\end{figure}

We study two different system geometries sketched in Fig. \ref{fig:geom}.

\emph{Subregion of an infinite system. }
The correlation matrix of an infinite translational invariant superconducting system can be written as
\begin{eqnarray}
 \bs{\chi}_{\bs r,\bs r'}& = & \int\frac{d^2 k}{(2\pi)^2} e^{i\bs k.\left(\bs r-\bs r'\right)}\bs{\chi}\left(\bs k\right)
\label{eq:chi_inf}
\end{eqnarray}
with $\bs{\chi}\left(\bs k\right)=\av{\bs C_{\bs k}\bs C_{\bs k}^{\dagger}}$ which, computed for a thermal Gibbs ensable with temperature $T=1/\beta$, equals $\bs{\chi}\left(\bs k\right) =  \bs 1-n_{f}\left[\bs{H}\left(\bs k\right)\right]$. For the ground-state result that will be used in the following one can simply take the $\beta\to\infty$ limit in which case $n_f$ becomes a Heaviside-theta function. Diagonalizing $\bs{H}\left(\bs k\right)$ explicitly we are thus able to compute $\bs{\chi}_A$ by numerically performing the integral in Eq.(\ref{eq:chi_inf}) for $r,r' \in A$.

\emph{Cylinder.  }
For a system placed  in a cylindrical geometry (see Fig. \ref{fig:geom}) the correlations matrix $ \bs \chi $ factorizes as a function of the momentum of the compact direction (here taken to be $k_x$):   
\begin{eqnarray}
\bs{\chi }_{\bs r,\bs r'} & = & \frac{1}{L_x} \sum_{k_{x}} e^{ik_x (x-x')} \bs{\chi }_{y,y'}\left(k_{x}\right)
\end{eqnarray}
where $\bs{\chi }\left(k_{x}\right) = \bs 1-n_{f}\left[\bs{H}\left( k_x \right)\right]$ with $\bs{H}\left( k_x \right)$ the $k_x$ component of the Hamiltonian Fourier-transformed in the $x$ direction and with open boundary conditions along the $y$ direction.
Further restricting this matrix to a subsystem $A$ that respects translational invariance along the $x$ direction, one may define
\begin{eqnarray}
\bs{\chi }_A(k_x) & = & \sum_{y,y' \in A} \ket{y} \bra{y} \bs{\chi }(k_x)  \ket{y'} \bra{y'}
\end{eqnarray}
 the $k_x$-resolved single particle correlation matrix in the $A$ domain.
In the same way the entanglement Hamiltonian can be $k_x$-resolved:  $\bs \Omega_A (k_x) = - \ln \left[ \bs{\chi }_A(k_x)^{-1}-1 \right]$. 
As a function of $k_x$ the entanglement entropy therefore factorizes  $ S\left[ \rho_{A} \right]  = \sum_{k_x}  S_{k_{x}}$ with 
\begin{eqnarray}
\label{eq:Skx}
S_{k_{x}} =  \sum_{\alpha}\mathcal{I}\left[\varepsilon_{\alpha}\left(k_{x}\right)\right]
\end{eqnarray}
 the  contribution of each momentum sector and where $\varepsilon_{\alpha}\left(k_{x}\right)$ are the eigenvalues of $\bs \Omega_A (k_x)$.

\section{Results}

The entanglement entropy, $S[\rho_A]$, computed for a square $L\times L$  subregion of size $L=6$ as a function of chemical potential and Zeeman term is given in  Fig. \ref{fig2} - left panel.
The transition lines of Fig.\ref{fig1} although visible are not particularly well defined for such sub-system size. 
Taking derivatives of the entanglement entropy with respect to the chemical potential or the Zeeman term, as shown in Fig.\ref{fig2} - (center and right panels), reveals that, even for such small sub-systems, the entanglement entropy 
clearly signals the transition lines and assumes rather different dependencies with respect to the 
parameters $\varepsilon_F$ and $M_z$, within the different phases. 
\begin{figure}[h]
\includegraphics[width=0.85\columnwidth]{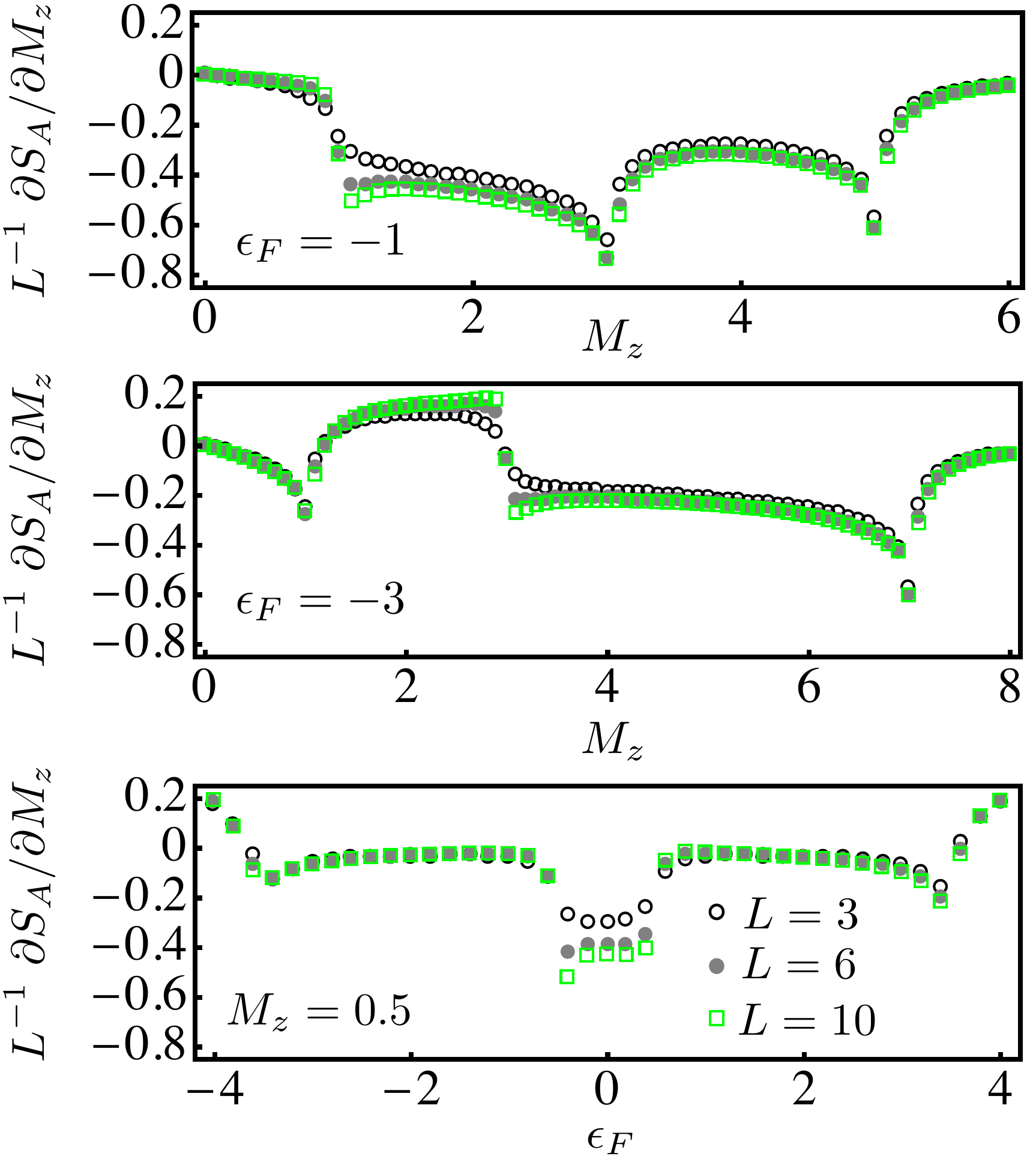}
\caption{\label{fig:S_A_derivatives}
Magnetization derivative of the local entanglement entropy of a square geometry with $L\times L$ sites for $L=3,6,10$ as a function of
Zeeman field $M_z$ for $\epsilon_F=-1,-3$  and as a function of the chemical potential $\epsilon_F$ for $M_z=0.5$. 
}
\end{figure}
For the same geometry, Fig.\ref{fig:S_A_derivatives} shows the convergence of the 
rescaled derivatives of the entanglement entropy with system size, along the three phase-space cuts 
of Fig.\ref{fig2}, for relatively small values of $L$.
Away from the phase transition lines, where the entanglement entropy clearly signals the transition, the derivative of the entanglement entropy is mildly varying within each phase displaying a plateau-like structure.
Indeed, inside each phase the derivatives reach a set of values that are approximately proportional to the value of the Chern number of the respective phase, with a proper rescaling of about $0.1$. 
Particularly, in the regime where the Chern number vanishes the entanglement entropy is fairly independent
of the magnetization.
Therefore, deep inside the phase,  the derivative of the entanglement entropy may be used to determine, with a good degree of accuracy, the Chern number of each topological phase. 
This has also been observed for other phase-space cuts. 
Even though in the superconducting phase a quantization is not expected, the plateau-like structure provides a
sensible signature of the phases. As shown before \cite{us3}, a similar result was obtained for the Hall conductivity
and its derivatives. Even though the Hall conductivity is also not quantized, it allows a clear signature of
the transitions between topological phases and, to some extent, provides information about the change in the Chern
number, although it does not provide information about the actual Chern number. In that respect, the entanglement
entropy provides a more detailed information about the topological phases.

In the following we study separately the extensive (proportional to the perimeter) and non-extensive 
contributions to the entanglement entropy, their dependence on the topological phase and on the 
geometry of the boundary.  We first consider the geometries displayed in the top panel of 
Fig.\ref{fig:shapes_and_xis} and compute the entanglement entropy using the methods explained in 
the previous sections. In the numerical calculations we typically consider systems of linear size $L \lesssim 30$.

\subsection{Entanglement entropy  - extensive contribution }

\begin{figure}[t]
\includegraphics[width=1.0\columnwidth]{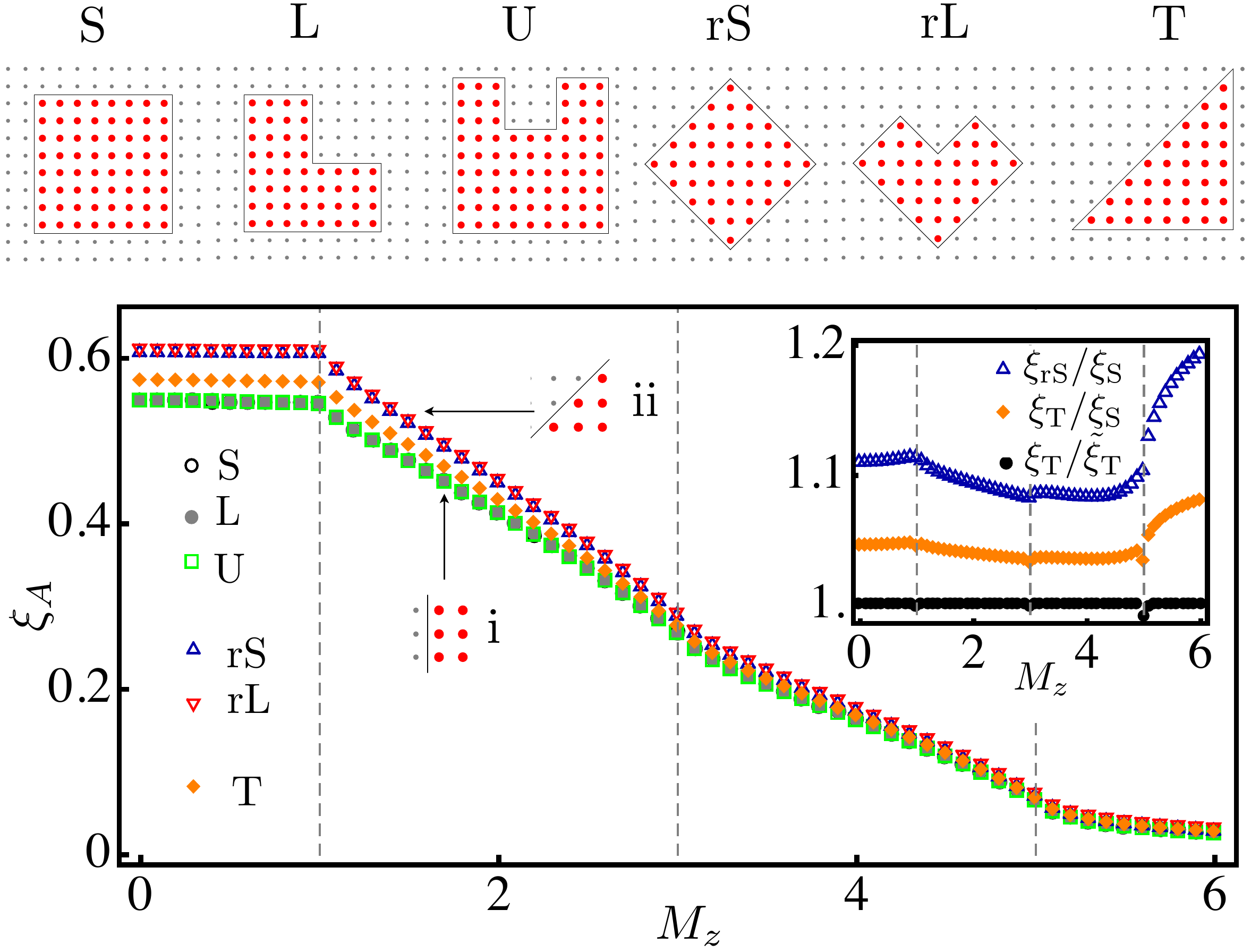}
\caption{\label{fig:shapes_and_xis}
Top panel: Different geometries of subsystems used for computing the entanglement entropy. The red dots correspond to the considered lattice degrees of freedom. The black lines define the boundary of a given geometry and were used to compute its perimeter $P$. Bottom panel: Coefficient of the perimeter contribution to the entanglement entropy, $\xi_A$, for different geometries, $A=\text{S}, \text{L}, ...$,  obtained fitting the numerically computed entanglement entropy for several system sizes and plotted as a function of $M_z$. Inset: Ratios of $\xi$'s as a function of $M_z$.  
}
\end{figure}

Fig.\ref{fig:shapes_and_xis}-(bottom panel) shows $\xi_A$ obtained by fitting the asymptotic large perimeter ($P$) limit for the different shapes along a cut in phase space where $\varepsilon_F=-1$ is kept constant. 
The square (S), L-shaped (L) and the U-shaped (U) yield the same $\xi_A$. 
The values of $\xi_A$  for the rotated square (rS) and the rotated L-shaped (rL) 
shapes also coincide but present a larger value than the one for S,L and U. 
This phenomena occurs as the two kinds of boundaries, labeled i and ii in the figure, have different orientations with respect to the underlying lattice.
The inset shows that the ratios $\xi_\text{rS}/\xi_\text{S}$ and  $\xi_\text{T}/\xi_\text{S}$  vary with $M_z$ and are therefore model specific. 

The triangular-like shape (T) has mixed boundaries, having a fraction $2/(2+\sqrt 2)$ of  type i boundary and a fraction $\sqrt 2 /(2+\sqrt 2)$ of type ii.
The values of $\xi_\text{T}$ lay between $\xi_\text{S}$ and $\xi_\text{rS}$. 
In the inset it is shown that the behaviour of  $\xi_\text{T}$ with $M_z$ can be reproduced defining $\tilde\xi_\text{T}=  ( 2 \xi_\text{T} + \sqrt{2} \xi_\text{rS} )/(2+\sqrt{2})$.

From these observations we conclude that $\xi_A$ varies with the specific details of the boundary and the ratio between different kinds of boundaries is not universal, depending on the specific details of the system. Nonetheless in systems with mixed boundaries their effect is additive and the total $\xi_A$ is an average over the $\xi$'s of the different types of boundaries.

\subsection{Entanglement entropy  - non-extensive contributions }

\begin{figure}[h]
\includegraphics[width=1.0\columnwidth]{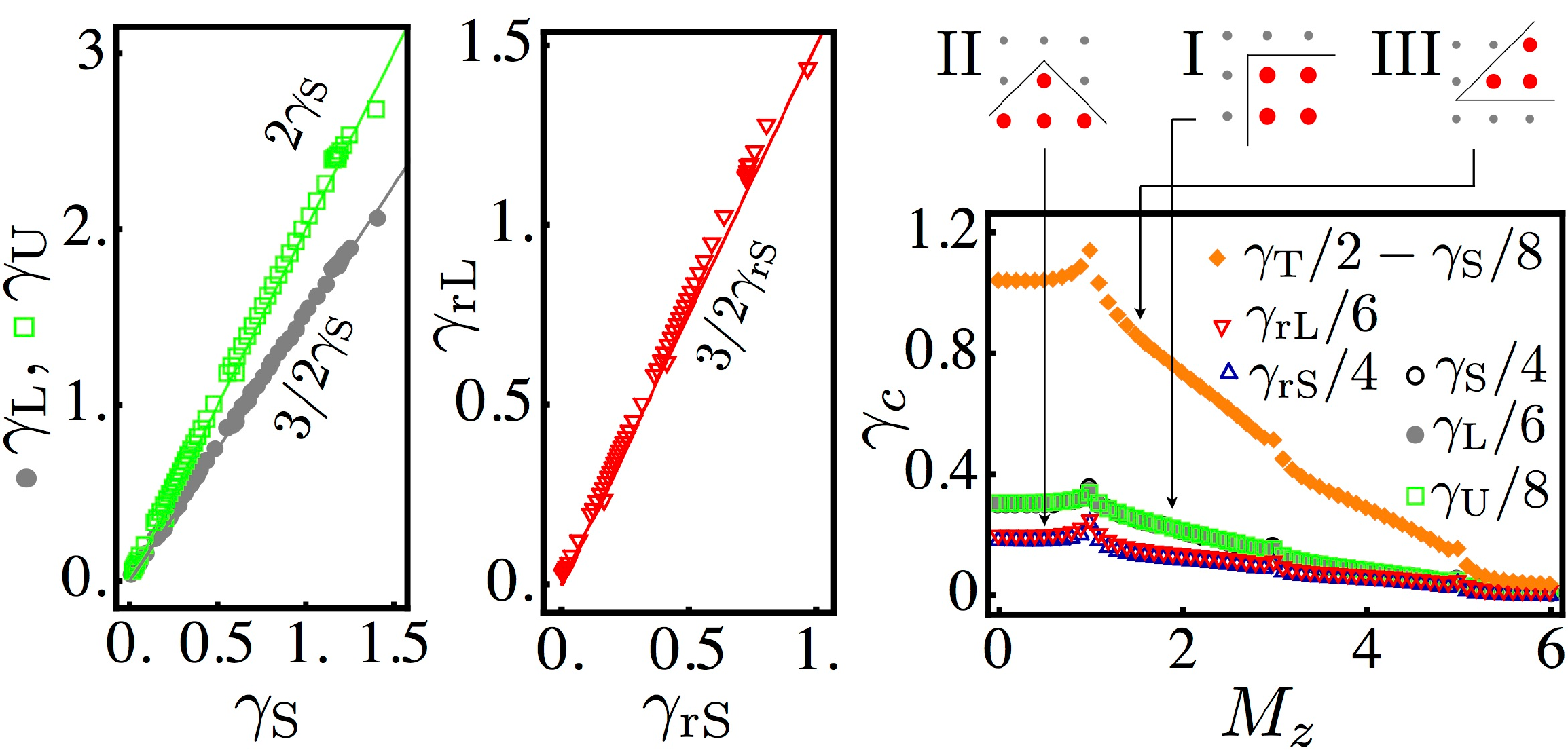}
\caption{\label{fig:gammas}
Left panel: Sub-leading correction for the geometries L and U as a function of the correction of the  S geometry. Middle panel:   Sub-leading correction for the geometries rL as a function of the rS one. Contribution of the different kinds of corners to the entanglement entropy.  
}
\end{figure}
\emph{Subregion of an infinite system. }
We now turn to the study of the non-extensive contributions. $\gamma_A$ for each geometry is obtained by fitting the sub-leading term of the entanglement entropy. 
 Fig.\ref{fig:gammas} -(left  panel) shows the fitted value of $\gamma_\text{L}$ and $\gamma_\text{U}$ as a function of $\gamma_\text{S}$ obtained by varying the Zeeman coupling $M_z$ at fixed $\varepsilon_F=-1$. This plot clearly shows that all the non-extensive contribution is due to the corners of the geometry. 
The  effect of  type I corners (see right panel) is additive as one obtains 
$\gamma_\text{S}/4=\gamma_\text{L}/6=\gamma_\text{U}/8$ i.e. $\gamma_A = n_A \gamma_I$ with $n_A$ 
the number of corners of the geometry. A similar conclusion can be deducted from  
Fig.\ref{fig:gammas} -(middle  panel) for the type II corners. 
The numerical verification of this fact enables us to compute the relative contribution of each kind of corner to the entanglement entropy. 
 Fig.\ref{fig:gammas} -(right  panel) shows the comparison of the different contributions as a function of $M_z$. It can be observed that, even if the difference of the contributions of type I and type II corners is mild, type III corners have a substantially higher contribution.  It was observed that the ratios $\gamma_\text{II}/\gamma_\text{I}$ and $\gamma_\text{III}/\gamma_\text{I}$ are dependent on the details of the system and are therefore non-universal quantities. A similar conclusion arrises when plotting $\gamma_A$ as a function of $\xi_A$.

\emph{ Vanishing of $\gamma_\text{Topo}$. }
\begin{figure}[h]
\includegraphics[width=0.99\columnwidth]{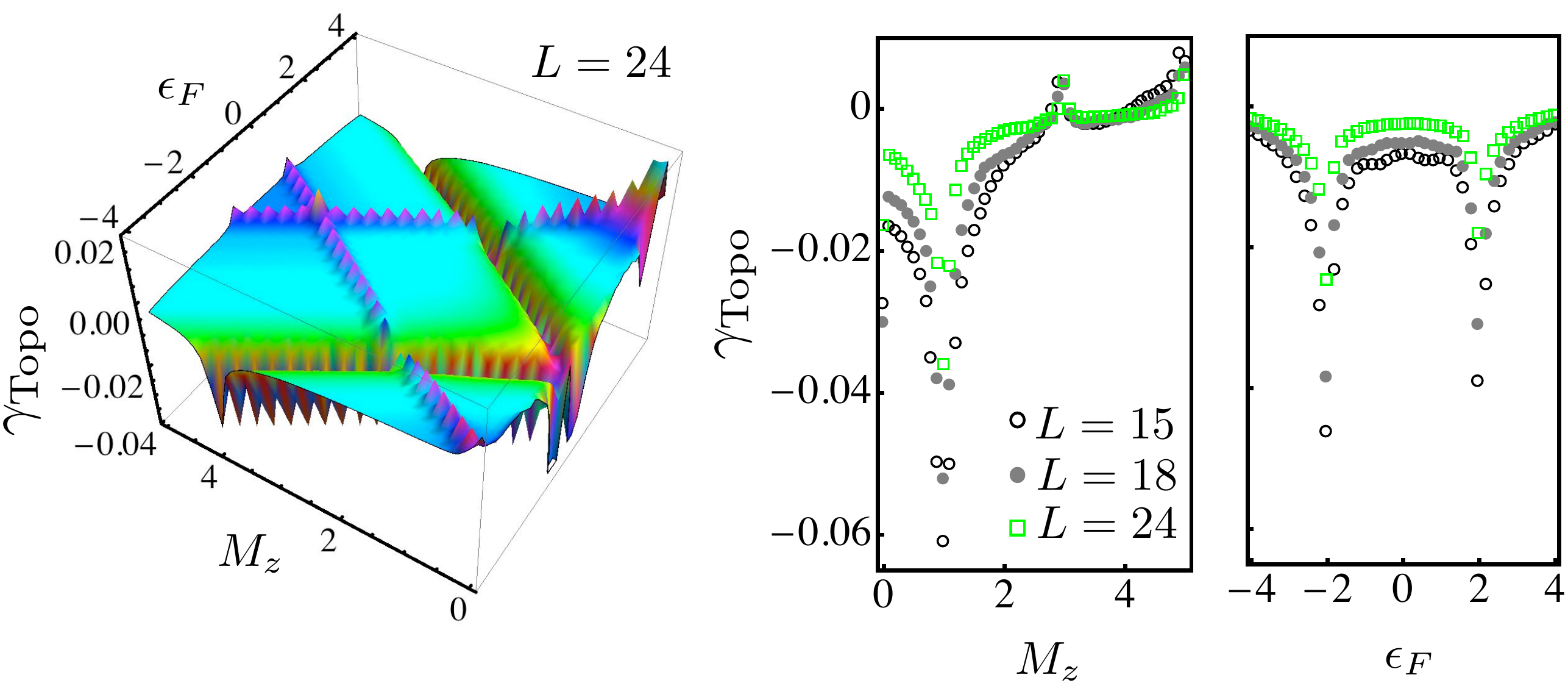}
\caption{\label{fig4}
Left panel: $\gamma_\text{Topo}$ for $L=24$ as a function of chemical potential and magnetization.
Middle and right panels: finite-size effects of $\gamma_\text{Topo}$ along cuts in the phase diagram.
}
\end{figure}
For a generic system, $\gamma_A$ may have contributions coming from the corners of the geometry as 
well as a topological one. For quadratic models the latter term is absent even if the system is in 
a non-trivial topological phase, as a topological insulator or superconductor.
Note that the relations $\gamma_\text{L}=3/2\gamma_\text{S}$ and  
$\gamma_\text{U}=2 \gamma_\text{S}$ obtained previously imply the absence of a 
topological term, since if present this would add a constant term to these relations. 
We now test the method described in Ref. \onlinecite{kitaev96} to compute the topological contribution, 
designed to minimise finite size effects. 
Here the so-called topological entanglement entropy is obtained considering three regions $A,B,C$ 
(defined in Fig.3 of Ref. \onlinecite{kitaev96})  
and calculating 
\begin{eqnarray}
\label{eq:Skx}
S_\text{Topo}  & =   &S_A+S_B+S_C-S_{AB}-S_{BC}-S_{AC}+S_{ABC} \nonumber \\
                       & =   & -2\gamma_\text{Topo}.
\end{eqnarray}
\begin{figure*}[t]
\includegraphics[width=0.9\textwidth]{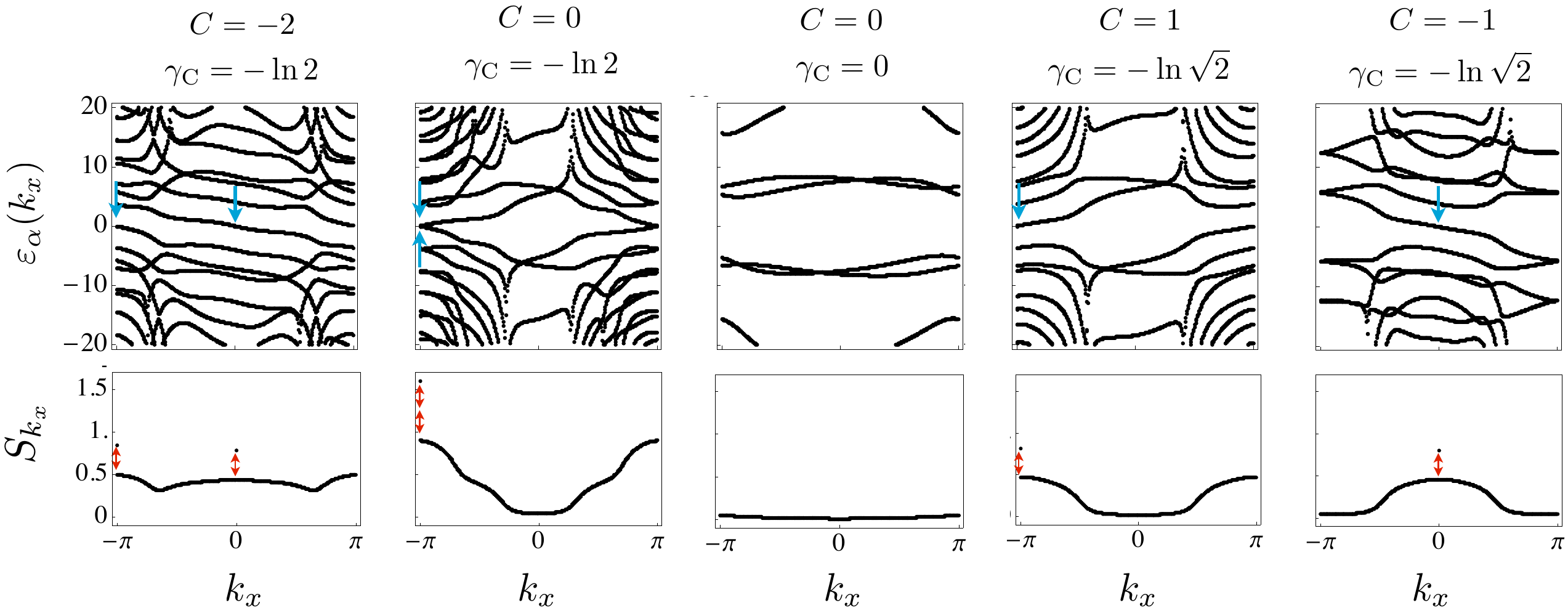}\caption{\label{fig:entanglement_spectum}
Upper panel: Single-particle entanglement
spectrum as a function of $k_{x}$ computed for a representative point of the different topological phases of Fig. \ref{fig1}, corresponding respectively to $\{\epsilon_F, M_z \} = \{1. ,2.\} ,\{ 2., 1.  \} ,\{ 6., 1. \} ,\{4., 2. \} ,\{ 2., 4. \} $ (from left to right) with $L_x = 400$ and $L_y = 70$. The blue arrows point to extra states arising at $\varepsilon_{\alpha}=0$ that yield the extra $1/2\ln2$
to the entropy. Lower panel: Contribution to the entanglement entropy
of each momentum sector. The red arrows correspond to an extra contribution
of $1/2\ln2$. }
\end{figure*}

In the first panel of Fig. (\ref{fig4}) we present results for $L=24$, as a function of the chemical potential and Zeeman term.
We consider the three regions $A,B,C$ immersed in an infinite system, for which we may calculate the correlations
matrix in terms of the momentum space solution of the topological superconductor.
The results clearly show the transition lines between the various phases. However, a close look into the
numerical results, shows that in some regions the topological entanglement entropy is positive but in some
it is negative. This is particularly aggravated for small system sizes. 
However, as shown in the other panels of Fig. (\ref{fig4}), far from the transition lines, $\gamma_\text{topo} \rightarrow 0$,
for all phases without distinction, as the system size grows. This is to be expected for a $\mathbb{Z}$ superconductor
(related to the integer quantum Hall effect for which there is no topological order, in the sense of Wen \cite{WenAP,yizhang}).
Therefore, in these geometries the TEE does not distinguish the various phases.

\subsection{ Cylinder geometry. }

We now address the entanglement entropy contribution on a cylindrical geometry (C) (see Fig. \ref{fig:geom}) following Ref. \onlinecite{balents}. For the numerical calculation we considered a cylinders up to $L_y=70$ and perimeter of varying sizes up to $L_x=400$. 

Concerning the  contribution  that is extensive with the perimeter of the system $P=L_x$, we find, as expected, a very good agreement between $\xi_S$ and $\xi_\text{C}$ as in this case the boundary is of type i (see Fig.\ref{fig:shapes_and_xis}).

As this geometry has no corners, the non-extensive component of the entanglement entropy, $\gamma_C$, receives contribution from edges states only. 
We find that in the topologically trivial phases \cite{sato,us3}, $C = 0$ (no edge states), $\gamma_C$ vanishes. In the topological phases we find that $\gamma_\text{C} \sim -\ln 2 $ when $C=0,-2$ (4 edge states) and $\gamma_\text{C} \sim - 1/2 \ln 2$ when $C=1,-1$ (2 edge states), as shown in Fig.\ref{fig1}. 
Thus, the contributions for the topological entanglement entropy can be written as $\gamma_\text{C} = -  (1/2 \ln 2)  g_{\text{edge}}$, where $g_{\text{edge}}$ is the number of edge state pairs.
Since the edge states are chiral, $\gamma_\text{C}$ seems to agree with
the $(1/2) \ln(2)$ found in chiral spin liquids \cite{yizhang}.
A similar result has been found for the Majorana mode of the n-channel Kondo model
with $n=2,S=1/2$, due to the impurity contribution \cite{Emery}.

Analysing each $\alpha$ contribution to the entanglement entropy, as given in Eq.(\ref{eq:S_0}), we conclude that
there is a reduced set of contributions that are independent of the system size. 
The remaining terms are not robust to changes in $L_x$. 
Their extrapolation to the infinite size limit gives a positive contribution to $\gamma_A$.
Since all contributions to the entropy are positive, and since the robust modes are invariant with the
system size these contribute a negative term to $\gamma_A$. As it turns out, this contribution has a larger
values than the positive one. 
This group of contributions give each a contribution of $(1/2) \ln(2)$ and are clearly related to the edge states.
We have checked that in a square geometry there are no robust modes (as the perimeter changes all $\lambda_\alpha$ values change as well).

In order to further understand these results we analyse in detail the entanglement spectrum for the cylinder geometry.
 Fig. (\ref{fig:entanglement_spectum}) shows the single-particle entanglement
spectrum for each of the phases of Fig. (\ref{fig1}) as a function
of $k_{x}$. Note that since the Hamiltonian is particle-hole symmetric
$\bs{\Omega}\left(k_{x}\right)=-\bs{\Omega}\left(-k_{x}\right)$.

When the entanglement spectrum is gapless the zero energy mode is doubly degenerate. 
This feature as been observed previously in a superconductor \cite{lukasz} and 
in the study of the entanglement spectrum of the Kitaev model \cite{Yao_2010}.
As a consequence, in the calculation of $S_{k_{x}}$, the degenerate states are responsible for an extra $1/2\ln2$ contribution: $S_{k_{x}}$ = $\tilde s(k_{x}) +1/2\ln2 \sum_p \delta_{k_x, k_p}  $, with $\tilde s(k) $ a continuous function of $k$ and where the  $k_p$'s ($ = 0, \pi$ in the examples of 
Fig. (\ref{fig:entanglement_spectum}) ) are the values of the momentum for which the degeneracy arises. 
The function $\tilde{s}\left(k_{x}\right)$ is plotted in Fig. (\ref{fig:entanglement_spectum}) 
where the $k_p$ points are also identified.  
For large values of $L_x$ the entanglement entropy can be approximated by 
\begin{eqnarray}
\label{eq:SrhoA}
S\left[\rho_{A}\right]  \simeq  \left[ \int\frac{dk_{x}}{2\pi}\tilde{s}\left(k_{x}\right) \right]  L_x + \left(1/2\ln2\right) g_{\text{edge}}
\end{eqnarray}
replacing the sum over momentum values by an integral. This corresponds to $ \xi_A = \int\frac{dk_{x}}{2\pi}\tilde{s}\left(k_{x}\right)$ and $\gamma_C=-\left(1/2\ln2\right)  g_{\text{edge}}$.

Fig. (\ref{fig:entanglement_spectum}) shows a striking difference between the entanglement 
spectrum of topological insulators and topologically trivial states: for $C \neq 0$, small 
deformations of the entanglement "bands" do not lead to a gapless entanglement spectrum.
 For $C=0$ two situations may arise: if there are no edge modes the entanglement spectrum is gapped 
(see Fig. (\ref{fig:entanglement_spectum}) third panel); in the presence of edge modes, it is still 
possible to open up a gap by lifting the degeneracy arising at $k_p$
(see Fig. (\ref{fig:entanglement_spectum}) second panel with $k_p=\pi$). 

In Ref. \onlinecite{Hughes_2010} a related but somehow different observation has been made by Hughes et al for the single-particle entanglement spectrum of Chern and spin Hall insulators. In their work the authors report that, for a trivial insulator with edge modes, even if the entanglement spectrum is gapless the bands that cross zero are disconnected from the rest of the spectrum.

\section{Discussion}

In this work we have analysed the entanglement entropy of a two-dimensional topological
p-wave superconductor. We have studied separately contributions to the entanglement entropy that are both extensive and non-extensive with the perimeter  of the subsystem. The main conclusions of our work are the following: 

i) The entanglement entropy clearly signals the topological transitions as its derivatives have sharp features around the transition lines, even for small systems sizes. Moreover, for our model, the derivative of the entanglement entropy with respect to the Zeeman term shows approximate plateaus that provide a  sensible signature of each topological phase.

ii) Due to the gapped nature of the spectrum away from the transitions between the various topological phases, the entropy obeys an area law with a non-universal pre-factor that depends on the parameters of the Hamiltonian but also on the specificities of the boundary of the subsystem. We find that boundary contributions coming from different kinds of boundaries simply add on. Having a way to estimate $\xi_A$ based on the characteristic lengths of the problem would be insightful, however in this study we found no clear relation of $\xi_A$ with the characteristic decay length of  various correlation functions.  

iii) Contributions to the entanglement entropy related to corners and edges of the boundary were found to be non-universal both with respect to the Hamiltonian coupling constants but also to the orientation of the corners with respect to the underlaying lattice. We found that corners with smaller angles give a larger contribution to the entanglement entropy. Our findings suggests that a systematic study of the dependence of corner contributions as a function of its angle could be an interesting avenue to further investigations.  

 iv) Even if the topological entanglement entropy  vanishes inside each topological phase, finite size effects are considerable and are quite severe in the proximity of transition lines, at least for the finite systems considered here. Our work suggests that an alternative way to find $\gamma_\text{Topo}$ is to study the relation between non-extensive corrections of different geometries as a parameter of the Hamiltonian is changed within the same topological phase. For example for systems with non-vanishing $\gamma_\text{Topo}$ the generic relation of $\gamma_\text{S}$ and $\gamma_\text{L}$ is $ 3 \gamma_\text{S} -2 \gamma_\text{L} =   \gamma_\text{Topo}$. Fitting such relation yields a better estimation for the TEE.   
 
 v) In a cylinder geometry the topological contribution to the entanglement entropy is finite and is in general
negative. The reason is associated with a robust negative contribution due to the
gapless edge states each contributing $(1/2)\ln 2$ to the entropy, characteristic
of Majorana edge states. The analysis of the entanglement spectrum shows that
the extra contributions to the entropy are associated with spectral degeneracies.

We thank discussions with V\'{\i}tor Rocha Vieira and partial
support by the Portuguese FCT under grant PEST-OE/FIS/UI0091/2011.

\end{document}